\documentclass[aps,twocolumn,preprintnumbers,nofootinbib,superscriptaddress]{revtex4-1}
\usepackage{amsmath} \usepackage{graphicx} \usepackage{amsfonts}
\usepackage{array} \usepackage{amsthm} \usepackage{bm}
\usepackage{palatino} \usepackage{mathpazo} 
\usepackage{supertabular}

\usepackage[breaklinks]{hyperref}
\usepackage{color}

\newcommand{\be}{\begin{equation}}
\newcommand{\ee}{\end{equation}}
\newcommand{\ba}{\begin{eqnarray}}
\newcommand{\ea}{\end{eqnarray}}
\newcommand{\bal}{\begin{align}}
\newcommand{\eal}{\end{align}}

\newcommand{\bw}{\begin{widetext}}
\newcommand{\ew}{\end{widetext}}

\begin{document}
\begin{flushleft}
\href{\doibase 10.1016/j.dark.2021.100802}{https://doi.org/10.1016/j.dark.2021.100802}
\end{flushleft}
\title{Rotating cosmological cylindrical wormholes in GR and TEGR sourced by anisotropic fluids}

\author{Mustapha Azreg-A\"{\i}nou}
\affiliation{Ba\c{s}kent University, Engineering Faculty, Ba\u{g}l\i ca Campus, 06790-Ankara, Turkey}


\begin{abstract}
Given an anisotropic fluid source, we determine in closed forms, upon solving the field equations of general relativity (GR) and teleparallel gravity (TEGR) coupled to a cosmological constant, cylindrically symmetric four-dimensional cosmological rotating wormholes, satisfying all local energy conditions, and cosmological rotating solutions with two axes of symmetry at finite proper distance. These solutions have the property that their angular velocity is proportional to the cosmological constant.
\end{abstract}


\maketitle

\section{Introduction\label{secI}}
{The teleparallel equivalent of general relativity (TEGR) is an equivalent formulation of Einstein's general relativity (GR), in that, the formulation ensures equivalence of the field equations as well as of the test particle equations of motion. For a review see the paper by Maluf~\cite{Maluf} where an account of the history of teleparallel theories of gravity is given. The two theories, TEGR and GR, use different connections, the curvature-less Weitzenb\"{o}ck connection~\cite{W} and the torsion-less Levi-Civita connection~\cite{LC}, respectively. The Weitzenb\"{o}ck connection has the property that it allows for a definition of a condition for absolute parallelism in space–time~\cite{Maluf}, hence the given name of ``teleparallel''. The tensors and invariants associated to TEGR exhibit no curvature but torsion only, that is, the information concerning the  gravitational field effects are encoded in the torsion tensor instead of the Riemann tensor, as it is the case in GR.}

{Despite their equivalence, the two theories are conceptually different. All physical features and results known in GR are also
described in the TEGR. The converse is not true: The TEGR approach allows the consideration of additional concepts and definitions. In the TEGR one meets tensors with three indices some of which are behind the definition of the gravitational energy–momentum which is consistent with the field equations~\cite{19,20}. The concept and definition of the gravitational angular momentum are also introduced in the TEGR~\cite{Maluf}. From the point of view of test particles, there is no notion of geodesics in the TEGR but only force equations and the source of force is torsion, more precisely the force is sum of a torsion-tensor component times two components of the velocity vector~\cite{book}. In the TEGR it is also possible to introduce the concept of inertia-free frames~\cite{book} and to split particle dynamical effects into distinct gravitational effects and purely inertial effects, that is, it is possible to separate inertia from gravity~\cite{nonzero}. Another example of conceptual difference between the two theories is provided in this work following Eq.~\eqref{pr}.}

{We have already mentioned some motivations for the investigation of the TEGR: Introduction of new concepts and unambiguous definitions such as that of the gravitational energy–momentum and other definitions~\cite{Maluf} and the clear distinction of purely inertial effects from gravitational ones. Other motivations for the TEGR is that the theory admits some extensions by adding quadratic and higher-order torsion terms to the action making it a good cosmological dark energy model without truly adding exotic matter to the cosmological field equations. Said otherwise, these extra terms added to the action, the so-called $f(T)$ gravity, have their counterparts in the cosmological field equations playing the ``role'' of exotic matter~\cite{Maluf,Maluf2}. The extended TEGR may provide a theoretical model to the late time universe acceleration problem~\cite{4}.}

Some rotating cylinders in general relativity sourced by anisotropic fluids and $\Lambda =0$ have been determined in~\cite{cz,lb,w4}. The purpose of this work is to consider the theories of GR and TEGR coupled to a cosmological constant and determine cylindrically symmetric rotating wormholes and other solutions sourced by anisotropic fluids.

Determining rotating and static solutions around an infinite axis is still a reviving topic. In GR there is a set of rotating cylindrically symmetric perfect fluid solutions which may be appropriate as matched interiors. Among the known solutions in GR we find the rotating dust of Vishveshwara and Winicour~\cite{fd1}, the perfect fluid sources with non-zero pressure of da Silva et al.~\cite{Lewis2}, Davidson~\cite{fd3,Davidson} and Ivanov~\cite{fd5}, and the family of Krasi\'{n}ski~\cite{fd6}. The solutions that will be constructed in this work share some physical and geometrical properties with the solutions known in the literature and have some other new properties. Among the new properties we mention that their angular velocity is proportional to the cosmological constant.

Investigating wormhole solutions is another, {rather many-fold}, reviving topic. {First of all, wormholes are special types of solutions to the field equations of gravity theories which contain two distant asymptotic regions or sheets, with a throat connecting the two and providing a shortcut for long journeys from one asymptotic region to another~\cite{MT,V}. Their observation has not been confirmed yet but they may be very well lurking in the universe. In Ref.~\cite{types} it was shown that the diameter of the shadow of type I supermassive wormhole overlaps with that of the black hole candidate at the center of the Milky Way. This shows that the existing up-to-date millimeter-wavelength very long baseline interferometry facilities do not lead to differentiate the supermassive black hole candidate at the center of the Milky Way from a possible type I supermassive wormhole.} Very recently, it was shown that the active galactic nuclei exhibit wormhole behaviors rather than supermassive black holes due to their gamma radiation resulting from collision of accreting flows~\cite{agn}. {Said otherwise, wormholes may be spotted in the sky upon detecting the gigantesque display of gamma rays that results from the collision of matter coming out of one mouth of the wormhole with infalling matter~\cite{agn}. A nearly similar conclusion was made in Ref.~\cite{types}: ``Other signals from the galaxy, as the motion of orbiting hot spots, may lead to draw a conclusion concerning the nature of the candidate''}.

{The interest to wormholes goes back to 1935. Einstein and N. Rosen were the first who described how two distant regions of spacetime can be joined together, creating a bridge between them. Most of the known solutions, and their list is too large to be cited here, are endowed with spherical symmetry, until recently a couple of wormhole solutions endowed with cylindrical symmetry were determined (see~\cite{symmetry} are references therein). The construction of analytical wormhole solutions, with spherical or cylindrical symmetry, provides the scientific community with theoretical tools for further investigations (calculations of shadow, quasi-normal modes, quasi-periodic oscillations, etc), as was done in~\cite{types,hot1,hot2}, and for computer simulations, as was done in~\cite{col1,col2} to study the collisional processes in the geometry of rotating wormholes.}

In Sec.~\ref{sectegr} we introduce the mathematical tools needed for the TEGR along with the necessary field equations. In Sec.~\ref{secT} we reduce the field equations. In Sec.~\ref{secEOS} we restrict ourselves to anisotropic fluids with $p_r=\omega_r \rho$, $p_\phi=\omega_\phi \rho$ and $p_z=\omega_z \rho$, where the equation-of-state (EoS) parameters ($\omega_r,\,\omega_\phi,\,\omega_z$) are constants constrained by $-1\leq \omega_r\leq 1$, $-1\leq \omega_\phi\leq 1$ and $-1\leq \omega_z\leq 1$. Section~\ref{secomega} is devoted to the construction of rotating wormholes and Sec.~\ref{rws} is devoted to the discussion of their physical and geometrical properties. In Sec.~\ref{sectas} we provide cosmological rotating solutions with two axes of symmetry at finite proper distance. Our final conclusions are given in Sec.~\ref{secc}. {An appendix section has been added to provide some useful formulas pertaining to Sec.~\ref{secomega}}.

{{\section{The teleparallel equivalent of general relativity\label{sectegr}}}
In the TEGR~\cite{HS9} the vielbein vector fields $e_a=e_{a}{^{\mu}}\partial_\mu$ are taken as fundamental variables instead of the metric $g_{\mu\nu}$, related to each other by
\begin{equation}\label{i1}
	g_{\mu \nu} =
	\eta_{ab}e^{a}{_{\mu}}e^{b}{_{\nu}},\quad g^{\mu \nu} =
	\eta^{ab}e_{a}{^{\mu}}e_{b}{^{\nu}},\quad e=\sqrt{|g|},
\end{equation}
with $\eta_{ab}=\text{diag}(+1,\,-1,\,-1,\,-1)$ being the metric of the 4-dimensional Minkowski spacetime and $e\equiv |\det(e^{a}{_{\mu}})|=\sqrt{|g|}$. In this work the tetrad indices, $a,\,b,\,c\cdots,k,\,l$, and Greek coordinate indices run from 0 to 3.

The field equation in teleparallel gravity in the presence of matter fields take the form~\cite{Bengochea,FF7}
\begin{align}\label{i2}
& I_\mu{^\nu}:=-\delta^\nu_\mu~\frac{f}{2} +2\Big[e^{-1}{e^a}_\mu\partial_\rho\Big(e{e_a}^\alpha
{S_\alpha}^{\rho \nu}\Big)-{T^\alpha}_{\lambda \mu}{S_\alpha}^{\nu \lambda}\Big]\nonumber\\
& =-\kappa T_{\text{(mat)}\,\mu}{}^\nu,
\end{align}
where $\kappa$ is the gravitational constant,
\begin{equation}\label{i3}
f(T)=T+2\Lambda ,
\end{equation}
and $T_{\text{(mat)}\,\mu}{}^\nu$ is the matter stress-energy tensor (SET) which we assume to be that of an anisotropic fluid of the form
\begin{align}\label{i4}
T_{\text{(mat)}\,\mu}{}^\nu & =\rho u_\mu u^\nu+\sum_{a=1}^3p_a{e_a}_\mu {e_a}^\nu\nonumber\\
& =(\rho+p_1)u_\mu u^\nu-p_1\delta^\nu_\mu+(p_2-p_1){e_2}_\mu {e_2}^\nu\nonumber\\
& +(p_3-p_1){e_3}_\mu {e_3}^\nu .
\end{align}
Here we have chosen ${e_0}^\nu=u^\nu$ to be the four-velocity vector of the fluid. The remaining quantities used in~\eqref{i2}, including the torsion $T$, are defined by\footnote{$S^{\alpha\mu\nu}$ may be given in a more compact form as:$$S^{\alpha\mu\nu}=\frac{1}{4}(T^{\nu\mu\alpha}+T^{\alpha\mu\alpha}-T^{\mu\nu\alpha})-\frac{1}{2}g^{\alpha\nu}{T^{\sigma\mu}}_{\sigma}+\frac{1}{2}g^{\alpha\mu}{T^{\sigma\nu}}_{\sigma}.$$}
\begin{align}\label{i5}
&{T^\alpha}_{\mu \nu}={e_b}^\alpha
(\partial_\mu{e^b}_\nu-\partial_\nu{e^b}_\mu),\nonumber\\
&K_{\alpha\mu\nu}=\frac{1}{2}~(T_{\mu\alpha\nu}+T_{\nu\alpha\mu}-T_{\alpha\mu\nu}),\nonumber\\
&S^{\alpha\mu\nu}=\frac{1}{2}~(K^{\mu\nu\alpha}-g^{\alpha\nu}{T^{\sigma\mu}}_{\sigma}+g^{\alpha\mu}{T^{\sigma\nu}}_{\sigma}),\nonumber\\
&T=T_{\alpha\mu\nu}S^{\alpha\mu\nu}.
\end{align}
In the definition of ${T^\alpha}_{\mu \nu}$ the connection $\omega^a{}_{b\mu}$ has been set equal to zero as this is always possible in teleparallel gravity~\cite{cov}.

In cylindrical coordinates ($x^0=t$, $x^1=r$, $x^2=\phi$, $x^3=z$), we introduce the following non-diagonal vielbein to describe rotating solutions
\begin{equation}\label{i6}
\big(e^{a}{_{\mu}}\big)= \begin{pmatrix}
{\rm e}^{\gamma (r)} & 0 & -{\rm e}^{-\gamma (r)} \Omega (r) & 0 \\
0 & {\rm e}^{\alpha (r)} & 0 & 0 \\
0 & 0 & {\rm e}^{\beta (r)} & 0 \\
0 & 0 & 0 & {\rm e}^{\mu (r)}
\end{pmatrix},
\end{equation}
resulting in the metric
\begin{multline}\label{m}
{\rm d}s^2=  {\rm e}^{2\gamma(r)}[ {\rm d}t - \Omega(r){\rm e}^{-2\gamma(r)}{\rm d}\phi ]^2- {\rm e}^{2\alpha(r)}{\rm d}r^2 \\- {\rm e}^{2\beta(r)}{\rm d}\phi^2
- {\rm e}^{2\mu(r)}{\rm d}z^2.\qquad\qquad\qquad\qquad
\end{multline}
Using all that in~\eqref{i5} we obtain
\begin{multline}\label{T}
T=\frac{1}{2} {\rm e}^{-2 (\alpha + \beta + \gamma )} (\Omega '-2 \Omega  \gamma ')^2\\
+2 {\rm e}^{-2 \alpha } (\beta ' \gamma '+\beta ' \mu '+\gamma ' \mu ').
\end{multline}
The nonvanishing components of the SET~\eqref{i4} are
\begin{align}\label{i4nv}
&T_{\text{(mat)}\,t}{}^t=\rho,\quad T_{\text{(mat)}\,r}{}^r=-p_r,\quad T_{\text{(mat)}\,\phi}{}^\phi=-p_\phi,\nonumber\\
&T_{\text{(mat)}\,z}{}^z=-p_z,\quad T_{\text{(mat)}\,\phi}{}^t=-\Omega (\rho+p_\phi){\rm e}^{-2 \gamma },
\end{align}
where we have set $p_1=p_r,\,p_2=p_\phi,\,p_3=p_z$.

It has become customary to introduce the vortex $\omega(r)$, which is the norm of the curl of the tetrad $e_{a}{^{\mu}}$~\cite{Bonnor} (see also~\cite{w1,w2,w3,w4}). This is related to $\Omega(r)$ by\footnote{The curl $\omega^\mu$ of $e_{a}{^{\mu}}$ is \[\omega^\mu:=\frac{1}{8}~\epsilon^{\mu\nu\rho\sigma}e_{a\nu}e^{a}{_{\rho;\sigma}},\] yielding $\omega^t=\omega^r=\omega^\phi=0$, $\omega^z=-{\rm e}^{\gamma-\alpha-\beta-\mu}(\Omega {\rm e}^{-2\gamma})'/2$ and $\omega=\sqrt{\omega^\mu\omega_\mu}={\rm e}^{\gamma-\alpha-\beta}(\Omega {\rm e}^{-2\gamma})'/2$.}
\begin{equation}\label{w}
\Omega (r):=2 {\rm e}^{2 \gamma (r)} \int^r {\rm e}^{\alpha (x)+\beta (x)-\gamma (x)} \omega (x){\rm d}x ,
\end{equation}
yielding
\begin{equation}\label{Ts}
T=2 \omega ^2+2 {\rm e}^{-2 \alpha } (\beta ' \gamma '+\beta ' \mu '+\gamma ' \mu ').
\end{equation}

One may bring the field equations~\eqref{i2} to\footnote{In this work {\[R^{\alpha}{}_{\beta\mu\nu}:=\partial_\nu\Gamma^{\alpha}_{\beta\mu}-\partial_\mu\Gamma^{\alpha}_{\beta\nu}+\Gamma^{\alpha}_{\eta\nu}\Gamma^{\eta}_{\beta\mu}-\Gamma^{\alpha}_{\eta\mu}\Gamma^{\eta}_{\beta\nu}.\]}}
\begin{equation}\label{eff}
G_{\mu}{}^\nu=-\kappa \tau_{\mu}{}^\nu
\end{equation}
where $G_{\mu}{}^\nu$ is the Einstein tensor and $\tau_{\mu}{}^\nu$ is the total SET including the cosmological constant and is defined by
\begin{align}\label{tau}
	&\tau_{t}{}^t=\rho-\frac{\Lambda}{\kappa },\quad \tau_{r}{}^r=-p_r-\frac{\Lambda}{\kappa },\nonumber\\
	&\tau_{\phi}{}^\phi=-p_{\phi }-\frac{\Lambda}{\kappa },\quad \tau_{z}{}^z=-p_z-\frac{\Lambda}{\kappa },\nonumber\\ 
	&\tau_{\phi}{}^t=-\Omega  (\rho +p_{\phi }){\rm e}^{-2 \gamma }.
\end{align}
Since $\nabla_\nu G_{\mu}{}^\nu\equiv 0$, one must have
\begin{equation}\label{must}
\nabla_\nu \tau_{\mu}{}^\nu= 0.
\end{equation}
The solutions we will derive in this work will satisfy the field equations of GR~\eqref{eff} and of TEGR~\eqref{i2}. For TEGR, the solutions are constrained by~\eqref{Ts}.

\section{Reducing the field equations\label{secT}}
Given that $T_{\text{(mat)}\,t}{}^\phi=0$ and $\tau_{t}{}^\phi=0$, the line $I_t{^\phi}=0$ in~\eqref{i2}, and the component  $_{t}{}^\phi$ of Eq.~\eqref{eff}, reduce upon using~\eqref{i4nv} and~\eqref{tau} to \[I_t{^\phi}=G_{t}{}^\phi=-{\rm e}^{\gamma- \alpha -\beta}[\omega(2\gamma'+\mu')+\omega']=0,\] yielding
\begin{equation}\label{omega}
\omega(r)=\omega_0{\rm e}^{-2\gamma -\mu},
\end{equation}
where $\omega_0$ is a constant of integration. Using this in~\eqref{w} we obtain
\begin{equation}\label{w2}
\Omega (r)=2\omega_0 {\rm e}^{2 \gamma (r)} \int^r {\rm e}^{\alpha (x)+\beta (x)-3\gamma (x)-\mu(x)} {\rm d}x .
\end{equation}

The expression of $G_{r}{}^r$ is just half that of $T$~\eqref{Ts}, $G_{r}{}^r=\omega ^2+{\rm e}^{-2 \alpha } (\beta ' \gamma '+\beta ' \mu '+\gamma ' \mu ')=T/2$, and this implies using~\eqref{eff} and~\eqref{tau} that $p_r$ is given by
\begin{equation}\label{pr}
p_r= \frac{T}{2 \kappa}-\frac{\Lambda}{\kappa} ,
\end{equation}
where the last term proportional to $\Lambda$ is the radial pressure due to the cosmological constant and the first term, $T/(2\kappa)$, is the radial pressure generated by a constant torsion. Thus, in the TEGR, a {\textquoteleft nonvanishing\textquoteright} torsion generates a nonvanishing radial pressure $T/(2\kappa)$ while in GR the relation~\eqref{pr} is written as $G_{r}{}^r=\kappa p_r+\Lambda$ and it merely expresses the fact that the geometric entity $G_{r}{}^r$ is proportional to the sum of the pressures $p_r$ and $\Lambda/\kappa$. {This is another example of conceptual difference between the two theories that we mentioned in the Introduction. This is at the level of the field equations, where the torsion appears as a force while its counterpart in GR, the curvature scalar $\mathcal{R}$, does not assume a dynamical role in the field equations.}

Substituting~\eqref{tau} into~\eqref{must} we obtain
\begin{equation}\label{ntau}
-4 p_r'-(\beta '+\gamma '+\mu ') p_r-\gamma ' \rho +\beta ' p_{\phi }+\mu ' p_z=0 .
\end{equation}

\section{Anisotropic fluids\label{secEOS}}
We consider the simple case where  $p_r=\omega_r \rho$, $p_\phi=\omega_\phi \rho$ and $p_z=\omega_z \rho$, with the EoS parameters ($\omega_r,\,\omega_\phi,\,\omega_z$) being constants generally constrained by
\begin{equation}\label{obey}
-1\leq \omega_r\leq 1,\quad -1\leq \omega_\phi\leq 1,\quad -1\leq \omega_z\leq 1 .
\end{equation}
The differential equation~\eqref{ntau} becomes
\begin{equation}\label{rho}
(\omega _r-\omega _{\phi }) \beta '+(1+\omega _r) \gamma '+(\omega _r-\omega _z) \mu '+4 \omega _r \frac{\rho
	'}{\rho }=0,
\end{equation}
resulting in
\begin{equation}\label{wr}
\rho =\rho _0\exp\Big[\frac{(\omega _{\phi }-\omega _r) \beta -(1+\omega _r) \gamma +(\omega _z-\omega _r) \mu}{4 \omega _r}\Big],
\end{equation}
where $\rho_0$ is a constant of integration. Using this in~\eqref{pr} we can evaluate the torsion from
\begin{equation}\label{torsion}
	T=2(\kappa\omega_{r}\rho+\Lambda).
\end{equation}

Since the radial coordinate $r$ can be changed at will by a coordinate transformation $r\to \bar{r}$, from now on we fix the coordinate gauge to be
\begin{equation}\label{gauge}
\alpha =\beta +\gamma +\mu .
\end{equation} 
In this gauge the independent field equations emanating from~\eqref{eff} 
\begin{equation}\label{R}
R_{\mu}{}^\nu=-\kappa [\tau_{\mu}{}^\nu -(1/2)\delta_{\mu}{}^\nu \tau_{\sigma}{}^\sigma],
\end{equation}
reduce to
\begin{align}
\label{eqs1}&2{\rm e}^{-2 \alpha } \gamma ''+4 \omega ^2=2\Lambda + \kappa  (1+\omega _r+\omega _{\phi }+\omega _z) \rho, \\
\label{eqs2}&2{\rm e}^{-2 \alpha } \mu ''=2\Lambda + \kappa  (-1+\omega _r+\omega _{\phi }-\omega _z) \rho, \\
\label{eqs3}&2{\rm e}^{-2 \alpha } \beta ''-4 \omega ^2=2\Lambda + \kappa  (-1+\omega _r-\omega _{\phi }+\omega _z) \rho, 
\end{align}
which are the $_{t}{}^t$, $_{z}{}^z$ and $_{\phi}{}^\phi$  equations~\eqref{R}, respectively. The $_{\phi}{}^t$ equation~\eqref{R} is a combination of~\eqref{eqs1} and~\eqref{eqs3} and the $_{r}{}^r$ equation~\eqref{R} is a combination of~\eqref{eqs1}, \eqref{eqs2}, \eqref{eqs3} and $G_{r}{}^r=\omega ^2+{\rm e}^{-2 \alpha } (\beta ' \gamma '+\beta ' \mu '+\gamma ' \mu ')$ [recall that $G_{r}{}^r=T/2$].

\section{Cosmological rotating wormholes\label{secomega}}
From now on, we assume $\omega_{r}\neq 0$ to ensure that~\eqref{wr} remains valid. We first construct cosmological rotating cylindrical wormholes to GR. The counterpart solutions to TEGR are the same with the torsion given by~\eqref{torsion}.

\begin{figure}[h]
\centering
  \includegraphics[width=0.49\textwidth]{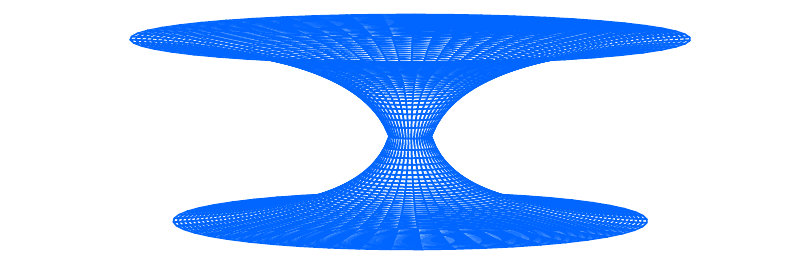} \\
  \caption{\footnotesize{The embedding diagram of the wormhole solution~\eqref{metric2} for $\delta=4/5$, $q=3/5$ and $u_0=1/10$.}}\label{Fig1}
\end{figure}
\begin{figure}[h]
	\centering
	\includegraphics[width=0.49\textwidth]{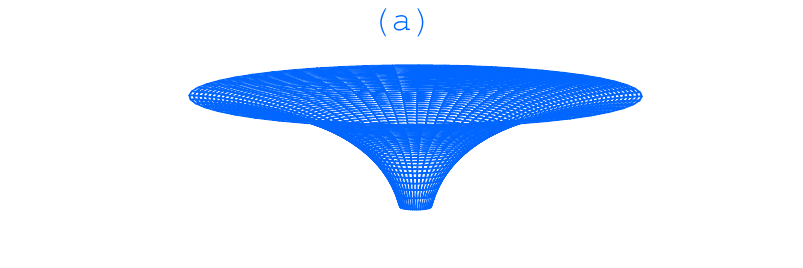} \includegraphics[width=0.49\textwidth]{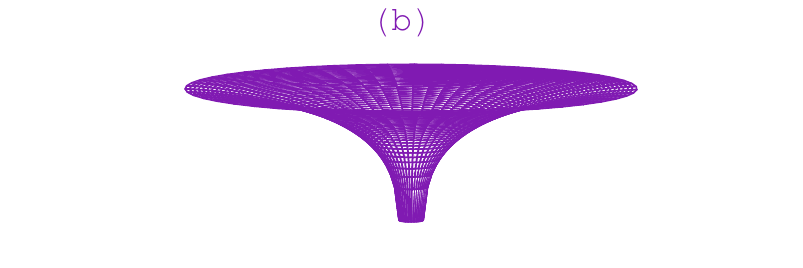} \includegraphics[width=0.49\textwidth]{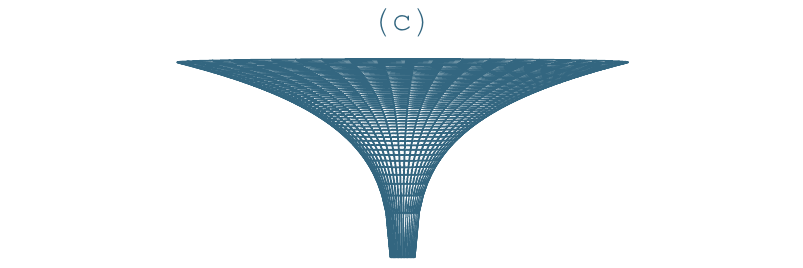} \\
	\caption{\footnotesize{The embedding diagram of the wormhole solution~\eqref{metric2} showing the upper sheet only for $u_0=1/10$. (a) $\delta=4/5$ and $q=3/5$ (upper panel), (b) $\delta=4/50$ and $q=3/5$ (intermediate panel), (c) $\delta=4/50$ and $q=3/50$ (lower panel).}}\label{Fig2}
\end{figure}

We look for solutions with a constant shift function, that is, $\gamma=0$ and $\mu=0$. In this case $\alpha=\beta$~\eqref{gauge} and Eq.~\eqref{eqs2} implies that $\rho$ is a constant given by $\rho_0=2\Lambda/[\kappa (1-\omega _r-\omega_{\phi }+\omega _z)]$. Since $\rho$ is constant, Eq.~\eqref{wr} implies
\begin{equation}\label{equal}
	\omega _{\phi }=\omega _r,
\end{equation}
and finally
\begin{equation}\label{rho1}
	\rho_0=\frac{2\Lambda}{\kappa (1-2\omega _r+\omega _z)}.
\end{equation}
Equations~\eqref{omega} and~\eqref{eqs1} imply that $\omega^2$ is also a constant proportional to the cosmological constant. This is given by
\begin{equation}\label{o02}
	\omega_0^2=\frac{\Lambda (1+\omega_{z})}{1-2\omega _r+\omega _z}=\frac{\kappa (1+\omega_{z})}{2}~\rho_0.
\end{equation}
Using these vales of ($	\rho_0,\,\omega_0^2$) in~\eqref{eqs3} we bring it to the form (recall $\alpha=\beta$)
\begin{equation}\label{form}
	{\rm e}^{-2 \alpha } \alpha ''=q^2\equiv \frac{2\Lambda (1-\omega _r+2\omega _z)}{1-2\omega _r+\omega _z}.
\end{equation}
This equation can be integrated in all three cases $q^2<0$, $q^2=0$ and $q^2>0$. In the case $q^2<0$ we obtain cosmological rotating solutions with two axes of symmetry at finite proper distance (see Sec.~\ref{sectas}). The case $q^2>0$ (we may assume $q>0$) yields cosmological rotating wormholes where the solution ${\rm e}^{2 \alpha }$  is brought to the form
\begin{equation}\label{alpha}
{\rm e}^{2 \alpha }={\rm e}^{2 \beta }=\frac{c^2\sec^2(c r)}{q^2},
\end{equation}  
with $c>0$ being a constant of integration. As to $\Omega$ is obtained from~\eqref{w2}
\begin{equation}\label{O}
	\Omega (r)=2\omega_0\int^r {\rm e}^{2\alpha (x)} {\rm d}x =\frac{2c\omega_0}{q^2}~\tan (cr),
\end{equation}
where we have dropped an additive constant of integration. The metric takes the form
\begin{multline}\label{metric}
	{\rm d}s^2=  \Big[ {\rm d}t -\frac{2c\omega_0\tan (cr)}{q^2}~{\rm d}\phi \Big]^2- \frac{c^2\sec^2(c r)}{q^2}~{\rm d}r^2 \\- \frac{c^2\sec^2(c r)}{q^2}~{\rm d}\phi^2
	- {\rm d}z^2.\qquad\qquad\qquad\qquad
\end{multline}
We see that the spherical radius ${\rm e}^{\beta }=c\sec(c r)/q$ has a minimum value at $r=0$ and increases as $|r|$ increases. On introducing the new radial coordinate $u=c\tan(cr)/q$ and the new constant $u_0^2=c^2/q^2$, we bring the metric~\eqref{metric} to the manifestly wormhole form
\begin{multline}\label{metric2}
	{\rm d}s^2=  \Big( {\rm d}t -\frac{2\omega_0 u}{q}~{\rm d}\phi \Big)^2- \frac{1}{q^2(u^2+u_0^2)}~{\rm d}u^2 \\- (u^2+u_0^2){\rm d}\phi^2
	- {\rm d}z^2.\qquad\qquad\qquad\qquad
\end{multline}
A very similar solution describing a rotating wormhole, which is a solution to the Einstein–Maxwell equations, was determined in~\cite{symmetry}.

The TEGR wormhole solution is also given by~\eqref{metric} and~\eqref{metric2} with $T=2\omega_0^2>0$~\eqref{torsion}, which is a positive constant.

The metric component
\begin{equation}\label{gff}
g_{\phi\phi}=-\Big(\frac{q^2-4\omega_0^2}{q^2}~u^2+u_0^2\Big),	
\end{equation}
is manifestly negative if $q^2-4\omega_0^2=2\Lambda(1+\omega_r)/(-1+2\omega_r-\omega_z)\geq 0$ signaling the absence of closed timelike curves (CTCs)~\cite{exact}. For  $q^2-4\omega_0^2< 0$, CTCs occur at large values of $|u|$.

Let us see under which conditions the constraints $q^2>0$, $\omega_0^2\geq 0$ and $\rho_0>0$ are satisfied simultaneously. The gravitational constant $\kappa$ being positive, we obtain the following constraints on  the EoS parameters ($\omega_r,\,\omega_\phi,\,\omega_z$) assuming they obey the general inequalities~\eqref{obey}. If $\Lambda<0$ we obtain
\begin{equation}\label{c1}
\frac{1}{3}<\omega_r\leq 1\;\text{ and }\;\frac{\omega_r-1}{2}<\omega_z<2\omega_r-1 .
\end{equation}
If $\Lambda>0$ we obtain
\begin{align}
\label{c2}&-1\leq \omega_r\leq \frac{1}{3}\;\text{ and }\;\frac{\omega_r-1}{2}<\omega_z\leq 1,\;\text{ or}\\
\label{c3}&\frac{1}{3}<\omega_r< 1\;\text{ and }\;2\omega_r-1<\omega_z\leq 1.
\end{align}
If, however, we want to avoid the presence of CTCs, we have to impose the fourth constraint $q^2-4\omega_0^2\geq 0$. This results in
\begin{equation}\label{c4}
\Lambda>0,\;\;\omega_r=-1\;\text{ and }\;-1< \omega_z\leq 1.
\end{equation}
This shows that, for a positive cosmological constant, it is always possible to have a rotating wormhole with a positive energy density and no CTCs. The anisotropic fluid is isotropic in a plane perpendicular to the axis of rotation with $\omega_{\phi }=\omega_r=-1$.

{To construct the embedding diagram of the metric~\eqref{metric2} we introduce the radial coordinate $R$ defined by $R^2=\delta u^2+u_0^2$ with $\delta\equiv (q^2-4\omega_0^2)/q^2$. For a constant time-slice and $z=\text{const.}$, the two-dimensional spatial metric takes the form \[	{\rm d}s_2^2=\frac{R^2{\rm d}R^2}{q^2(R^2-u_0^2)[R^2-(1-\delta)u_0^2]}+R^2	{\rm d}\phi^2.\] This is to be embedded in the three-dimensional
Euclidean space \[{\rm d}s_3^2={\rm d}R^2++R^2	{\rm d}\phi^2+{\rm d}Z^2.\] The embedded diagram is a surface of revolution, symmetric with respect to the plane $Z=0$, where $Z\equiv Z(R)$ with \[Z(R)=\pm\int_{u_0\sqrt{1+\delta}}^R\sqrt{\frac{x^2}{q^2(x^2-u_0^2)[x^2-(1-\delta)u_0^2]}-1}~{\rm d}x.\] Plots of the surface of revolution as depicted in Figs.~\ref{Fig1} and~\ref{Fig2}. In Fig.~\ref{Fig1} we have shown the two sheets of the wormhole for some set of the parameters ($\delta,\,q,\,u_0$) while in Fig.~\ref{Fig2} only the upper sheet is depicted for different values of the parameters ($\delta,\,q,\,u_0$).}

\section{Physical and geometrical properties of the rotating wormhole solution\label{rws}}
Since the wormhole solutions~\eqref{metric} and~\eqref{metric2} are not asymptotically flat, they are appropriate as matched interiors. Another reason why they are so is that the speed of rotation, $\Omega =2\omega_0 u/q$~\eqref{O}, increases linearly as one moves away from the axis of rotation at $u=0$. Consequently the linear speed of each fluid particle increases unceasingly as one moves away from the axis of rotation. Following the work done in~\cite{symmetry} one can match these wormholes to external rotating flat Minkowskian metrics in such a way that the radii of the cylindrical junction surfaces are chosen so that the interior wormhole region is exempt from CTCs. 

Thus, the constraints~\eqref{c1}-\eqref{c3}, for $\Lambda<0$ and $\Lambda>0$, are fairly enough to obtain physically acceptable wormhole solutions that certainly do not violate the weak energy condition ($\rho_0\geq 0$, $\rho_0+p_r\geq 0$, $\rho_0+p_z\geq 0$) and the dominant energy condition ($\rho_0\geq 0$, $-\rho_0\leq p_r\leq \rho_0$, $-\rho_0\leq p_z\leq \rho_0$). To satisfy the requirements of the strong energy condition ($\rho_0\geq 0$, $\rho_0+p_r\geq 0$, $\rho_0+p_z\geq 0$, $\rho_0+2p_r+p_z\geq 0$), the EoS parameters must obey the constraints~\eqref{c1} if $\Lambda<0$ and the following constraints if $\Lambda>0$:
\begin{align}
\label{c5}&\omega_r=-1\;\text{ and }\;\omega_z=-1,\;\text{ or}\\
\label{c6}&-1< \omega_r< -\frac{1}{5}\;\text{ and }\;-2\omega_r-1\leq\omega_z\leq 1,\;\text{ or}\\
\label{c7}&-\frac{1}{5}\leq\omega_r\leq\frac{1}{3}\;\text{ and }\;\frac{\omega_r-1}{2}<\omega_z\leq 1,\;\text{ or}\\
\label{c8}&\frac{1}{3}<\omega_r<1\;\text{ and }\;2\omega_r-1<\omega_z\leq 1.
\end{align}

Considered as interiors all the matter should be distributed inside cylindrical surfaces $\Sigma_+$ and $\Sigma_-$ of finite radii $u_+>u_0$ and $u_-<-u_0$. Referring to~\cite{Bonnor,Whittaker}, the effective mass $m_+$ and angular momentum $j_+$ per unit $z$-coordinate length of matter enclosed by $\Sigma_+$ are defined by (there are similar definitions for $m_-$ and $j_-$ concerning the matter distribution enclosed by $\Sigma_-$):
\begin{align}
&m_+=2\pi \int_{0}^{u_+}\tau_{t}{}^\mu n_\mu\sqrt{|g|}~{\rm d}u,\\
&j_+=-2\pi \int_{0}^{u_+}\tau_{\phi}{}^\mu n_\mu\sqrt{|g|}~{\rm d}u,
\end{align}
where $\tau_{\nu}{}^\mu$ is the total SET given in~\eqref{tau} and $g$ is the determinant of the metric. The vector $n_\mu$ is the unit normal to the spacelike surface of integration, which is the hypersurface $t=\text{const.}$ yielding $n_\mu=\delta_{\mu}^t/\sqrt{g^{tt}}$ with $g^{tt}$ being the component $tt$ of the inverse metric. Using~\eqref{metric2} along with $g=-1/q^2$ we find
\begin{align}
\label{mp}&m_+=\frac{2\pi u_0}{q} \Big(\rho_0-\frac{\Lambda}{\kappa}\Big) \int_{0}^{x_+}\sqrt{\frac{x^2+1}{\delta x^2+1}}~{\rm d}x,\\
\label{jp}&j_+=\frac{4\pi u_0^2\rho_0\omega_0(1+\omega_r)}{q^2} \int_{0}^{x_+}x\sqrt{\frac{x^2+1}{\delta x^2+1}}~{\rm d}x,
\end{align}
where $p_\phi=p_r=\omega_r \rho_0$ has been used, $x\equiv u/u_0$ and $x_+\equiv u_+/u_0$. These integrals {are expressed in terms of the incomplete elliptic integral of the second kind and the log function as given in the Appendix}. It is obvious that $j_+$ has the sign of $\omega_0$ and is zero in the static case (no rotation) and in the extreme case $\omega_r=-1$. The cosmological constant being small, we expect that in physically interesting situations to have $\rho_0>|\Lambda|/\kappa$ and thus $m_+>0$.

The constants of integration ($\omega_0,\,u_0/q$) and $\rho_0\propto \omega_0^2$~\eqref{o02} {could be determined analytically in terms of ($m_+,\,j_+,\,x_+$) if the expressions of ($m_+,\,j_+$), as given in the Appendix, were inversible. This is, however, possible for $x_+\ll 1$ where ($m_+,\,j_+$) expand as (for all $\delta$):
\begin{align*}
&m_+=\frac{\pi  u_0 (\kappa  \rho _0-\Lambda ) x_+ [6+(1-\delta ) x_+^2]}{3 q \kappa }+\mathcal{O}(x_+^5),\\
&j_+=\frac{\pi  u_0^2 \rho _0 \omega _0 (1+\omega _r) x_+^2 [4+(1-\delta ) x_+^2]}{2 q^2}+\mathcal{O}(x_+^6).
\end{align*}
On eliminating the ratio $u_0/q$ and using $\rho_0=2\omega_0^2/[\kappa(1+\omega_{z})]$~\eqref{o02}, we arrive at
\begin{multline*}
9 \kappa  m_+^2  (1+\omega _z) (1+\omega _r)[4+(1-\delta ) x_+^2] \omega _0^3\\ = \pi  j_+ [6+(1-\delta ) x_+^2]^2
[\Lambda  (1+\omega _z)-2 \omega _0^2]^2,
\end{multline*}
which can be solved by radicals for $\omega_0$ in terms of ($m_+,\,j_+,\,x_+$). The roots of this quartic equation in $\omega_0$, which can be obtained using a computer algebra system, are sizable and cannot be reproduced in this work. Then, the ratio $u_0/q$ is obtained from the above-given expansion of $m_+$. Finally, using the expression of $q$~\eqref{form} we can determine all the constants of integration in terms of the physical quantities ($m_+,\,j_+,\,x_+,\,\omega_{r},\,\omega_{z},\,\Lambda,\,\kappa$)}.

The curvature scalar, $\mathcal{R}$, and the Kretschmann scalar, $R^{\mu\nu\alpha\beta}R_{\mu\nu\alpha\beta}$, assume the following expressions:
\begin{equation}\label{scalar}
\mathcal{R}=2 (\omega_0^2-q^2),
\end{equation}
\begin{align}
&R^{\mu\nu\alpha\beta}R_{\mu\nu\alpha\beta}=4(q^4-6\omega_0^2q^2+11\omega_0^4),
\end{align}
which are finite constants.

\section{Cosmological rotating solutions with two axes of symmetry at finite proper distance\label{sectas}}
If the constant $q^2$ in~\eqref{form} is negative we set $Q^2=-q^2$ in~\eqref{form} and the solution yields
\begin{equation}\label{alphan}
	{\rm e}^{2 \alpha }={\rm e}^{2 \beta }=\frac{c^2\text{sech}^2(c r)}{Q^2},\quad \Omega =\frac{2c\omega_0}{Q^2}~\tanh (cr),
\end{equation}  
where we have dropped an additive constant $\Omega_0$ in the expression of $\Omega$. On setting $u=c\tanh(cr)/Q$ and $u_0^2=c^2/Q^2$, we bring the metric to the form
\begin{multline}\label{metricQ}
	{\rm d}s^2=  \Big[ {\rm d}t -\frac{2c\omega_0\tanh (cr)}{Q^2}~{\rm d}\phi \Big]^2- \frac{c^2\text{sech}^2(c r)}{Q^2}~{\rm d}r^2 \\- \frac{c^2\text{sech}^2(c r)}{Q^2}~{\rm d}\phi^2
	- {\rm d}z^2,\qquad\qquad\qquad\qquad
\end{multline}
\begin{multline}\label{metric2Q}
	{\rm d}s^2=  \Big( {\rm d}t -\frac{2\omega_0 u}{Q}~{\rm d}\phi \Big)^2- \frac{1}{Q^2(u_0^2-u^2)}~{\rm d}u^2 \\- (u_0^2-u^2){\rm d}\phi^2
	- {\rm d}z^2.\qquad\qquad\qquad\qquad
\end{multline}
Note that this solution shares with the wormhole solution~\eqref{metric}-\eqref{metric2} the same values of the physical quantities ($\rho_0,\;\omega_0^2$), given in~\eqref{rho1} and~\eqref{o02}. The corresponding TEGR solution is also given by~\eqref{metricQ}-\eqref{metric2Q} and shares the same value of the torsion $T=2\omega_0^2$ with the wormhole solution~\eqref{metric}-\eqref{metric2}.

The circular radius ${\rm e}^{\beta}=\sqrt{u_0^2-u^2}$ vanishes at $u=\pm u_0$ signaling the presence of two axes of symmetry. In the vicinity of these two axes the solution~\eqref{metricQ}-\eqref{metric2Q} has CTCs where $g_{\phi\phi}=-(u_0^2-u^2-4\omega_0^2u^2/Q^2)$ becomes positive (the nonrotating solution, $\omega_0=0$, has no CTCs). The integral \[\int_{-u_0}^{u_0}\frac{1}{Q\sqrt{u_0^2-u^2}}~{\rm d}u=\frac{\pi}{Q},\] being convergent, the two axes are at finite proper distance from each other and there is no spatial infinity. The absence of spatial infinity is also known for the nonrotating Melvin solution~\cite{exact,exact2,Melvin}, however, for the latter the proper distance of the two axes of symmetry \[\int_{0}^{\infty}(1+B^2u^2){\rm d}u,\]
diverges ($B$ is the magnetic field). Such nontrivial behaviors of the intrinsic geometry are familiar with static and rotating, cylindrically symmetric and/or axially symmetric, metrics and more examples are provided in~\cite{exact}.

\section{Conclusion\label{secc}}
As we mentioned in the Introduction, the determination of rotating and static solutions around an infinite axis is still attracting much attention. In this work, we presented a first set of two cosmological (energy density constant) rotating solutions in GR and TEGR gravity sourced by anisotropic fluids (isotropic in a plane perpendicular to the axis of symmetry) and extended the existing list of solutions pertaining to GR. These solutions have the property that their angular velocity is proportional to the cosmological constant.

We have shown that the EoS parameters obey a large set of values ensuring the satisfaction of all local energy conditions for the rotating wormholes. Such solutions can straightforwardly be matched to exterior rotating Minkowskian metrics. 

The other cosmological rotating solution has two axes of symmetry at finite proper distance where one axis can be regularized upon appropriately fixing the value of the additive constant of integration in the expression of $\Omega$ (which we have dropped) at the expense of rendering the time coordinate periodic.



%

\section*{Appendix: Analytical expression for ($\pmb{m_+,\,j_+}$)\label{seca}}
\renewcommand{\theequation}{A.\arabic{equation}}
\setcounter{equation}{0}
{The analytical expressions of ($m_+,\,j_+$) depend on the sign of $\delta$, however, their series expansions about $x_+=0$ do not, as this is shown in the new paragraph following~\eqref{jp}. For $\delta>0$ we have
\begin{align*}
&m_+=-\frac{2 {\rm i} \pi  u_0 (\kappa  \rho _0-\Lambda )}{\kappa  q \sqrt{\delta }}~ E\Big({\rm i} \ln (\sqrt{\delta } x_++\sqrt{1+\delta 
	x_+^2})\big|\frac{1}{\delta }\Big),\\
&j_+=\frac{2 \pi  u_0^2 \rho _0 \omega _0 (1+\omega _r)}{\delta^{3/2}  q^2}\bigg[\sqrt{1+\delta  x_+^2} \sqrt{\delta  (1+x_+^2)}-\sqrt{\delta }\\
&+(1-\delta )  \ln \bigg(\frac{1+\sqrt{\delta
}}{\sqrt{1+\delta  x_+^2}+\sqrt{\delta  (1+x_+^2)}}\bigg)\bigg].
\end{align*}
For $\delta=0$, we have
\begin{align*}
&m_+=\frac{\pi  u_0 (\kappa  \rho _0-\Lambda ) [x_+ \sqrt{1+x_+^2}+\ln (x_++\sqrt{1+x_+^2})]}{\kappa  q},\\
&j_+=\frac{4 \pi  u_0^2 \rho _0 \omega _0 (1+\omega _r) [(1+x_+^2) \sqrt{1+x_+^2}-1]}{3 q^2}.
\end{align*}
For $\delta<0$ we obtain
\begin{align*}
&m_+=\frac{2 \pi  u_0 (\kappa  \rho _0-\Lambda )}{\kappa  q \sqrt{-\delta }}~ E\Big(-{\rm i} \ln ({\rm i}\sqrt{-\delta } x_++\sqrt{1+\delta 
		x_+^2})\big|\frac{1}{\delta }\Big),\\
&j_+=\frac{2 \pi  u_0^2 \rho _0 \omega _0 (1+\omega _r)}{\delta^{3/2}  q^2}\bigg[{\rm i}\sqrt{1+\delta  x_+^2} \sqrt{-\delta  (1+x_+^2)}-{\rm i}\sqrt{\delta }\\
&+(1-\delta )  \ln \bigg(\frac{1+{\rm i}\sqrt{-\delta
}}{\sqrt{1+\delta  x_+^2}+{\rm i}\sqrt{-\delta  (1+x_+^2)}}\bigg)\bigg],
\end{align*}
where $0<x_+<1/\sqrt{-\delta}$ if $\delta<0$}.

{In all these expressions ${\rm i}^2=-1$ and \[E(z |m)\equiv\int_{0}^{z}\sqrt{1-m\sin^2\theta}~{\rm d}\theta ,\] is the incomplete elliptic integral of the second kind. Here $z$ is generally a complex number input. Despite a complex input, the output of the above-given expressions of ($m_+,\,j_+$) is always real for all $\delta$ (for $\delta<0$ the output is real provided $0<x_+<1/\sqrt{-\delta}$).}


\end{document}